\documentclass[hys]{svmult}

\usepackage{makeidx}         
\usepackage{graphicx}        
\usepackage{latexsym}                             
\usepackage{multicol}        
\usepackage[bottom]{footmisc}

\makeindex             

\begin{document}

\title*{Fractal Dimension\index{fractal dimension} of the El Salvador\index{El Salvador} Earthquake\index{earthquake}(2001) time Series}
\author{Fractal Dimension\index{fractal dimension} of the El Salvador\index{El Salvador} Earthquake\index{earthquake}(2001) time Series}
\author{Md. Nurujjaman
\and Ramesh Narayanan
\and A.N.Sekar Iyengar}
\institute{Saha Institute of Nuclear Physics, 1/AF, Bidhannagar, Kolkata-700 064, India }
%
\maketitle
\section{Introduction}
Earthquakes \index{earthquake} occur on the earth's surface as a  result of rearrangement of terrestrial cortex or higher part of the mantle. The energy released in this process propagates over long distances in the form of elastic seismic waves \cite{jaman:Gafarov}. In order to predict earthquakes \index{earthquake} many models have been proposed \cite{jaman:pradhan,jaman:burridge}. Dynamics of an earthquake\index{earthquake} is so complicated that it is quite difficult to predict using available models.  
Seismicity is a classic example of a complex phenomenon that can be quantified using fractal concepts \cite{jaman:Nanjo}.

In this paper we have estimated the fractal dimension\index{fractal dimension}, maximum, as well as minimum of the singularities, and the half-width of the multifractal spectrum of the El Salvador\index{El Salvador} Earthquake\index{earthquake} signal at different stations. The data has been taken from the California catalogue (http://nsmp.wr.usgs.gov/nsmn$\_$ eqdata.html). The paper has been arranged as follows: In section~\ref{sec:multifractal} the basic theory of  multifractality has been discussed, and the results have been presented in section~\ref{ref:Result} .
    
\section{ Multifractal Analysis}
\label{sec:multifractal}
The H\"older exponent\index{H\"older exponent} of a time series $f(t)$ at the point $t_0$ is given by the largest exponent such that there exists a polynomial $P_n(t-t_0)$ of the order of $n$ satisfying \cite{jaman:Muzy1,jaman:Bacry,jaman:Muzy2}

\begin{equation}
\label{jaman:eqn1}
|f(t)-p_n(t-t_0)|\leq C|t-t_0|^{\alpha(t_0)}
\end{equation}

The polynomial $P_n(t-t_0)$ corresponds to the Taylor series of $f(t)$ around $t=t_0$, up to $n$. The exponent $\alpha$ measures the irregularities of the function $f$. Higher positive value of $\alpha(t_0)$ indicates regularity in the function $f$. Negative $\alpha$ indicates spike in the signal. If $n<\alpha<n+1$ it can be proved that the function $f$ is $n$ times differentiable but not $n+1$ times at the point $t_0$ \cite{jaman:Arneodo}.

All the H\"older exponent\index{H\"older exponent}s present in the time series are given by the singularity spectrum D($\alpha$). This can be determined from the  Wavelet Transform Modulus Maxima\index{Wavelet Transform Modulus Maxima}(WTMM). Before proceeding to find out the exponents $\alpha$ using wavelet analysis, we discuss about the wavelet transform.
\subsection{Wavelet Analysis}
\begin{figure}[h]
\center
\includegraphics[height=4cm]{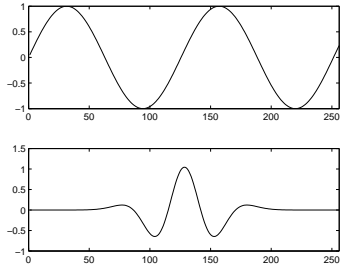}
\caption{A sinusoidal signal(up), and the gaussian Wavelet having four vanishing moments(bottom). Sinusoid has no limitation of duration where as wavelet has a limited duration.} 
\label{jaman:fig1}
\end{figure}

In order to understand wavelet analysis, we have to first understand `\emph{What is a wavelet}?'. A wavelet is a waveform of effectively limited duration that has an average value of zero, shown in the figure~\ref{jaman:fig1}~(bottom). The difference of wavelets to sine waves, which are the basis of Fourier analysis, is that sinusoids do not have limited duration, but
they extend from minus to plus infinity. And where sinusoids are smooth and predictable, wavelets tend to be irregular and asymmetric. For example, `gaus4' wavelet (Fig.~\ref{jaman:fig1}(bottom)) is defined as $\psi(t)=\frac{d^N}{dt^N}e^{-t^2/2}$, where $N$=4. Fourier analysis breaks up a signal into sine waves of various frequencies. Similarly, wavelet analysis breaks up a signal into shifted and scaled versions of the original (or mother) wavelet. It can be intuitively understood that signals with sharp changes might be better analyzed with an irregular wavelet than with a smooth sinusoid. Local features can be described better with wavelets that have local extent.

Wavelet transform can be defined as 
\begin{equation}
\label{jaman:eqn2}
Wf(s,b)=\frac{1}{s}\int_{-\infty}^{+\infty} f(t)\psi(\frac{x-b}{s})dt
\end{equation} 
where $s$, and $b$ are the scale and time respectively. In order to detect singularities we will further require $\psi$ to be orthogonal to some low-order polynomials \cite{jaman:Arneodo}:
\begin{equation}
\label{jaman:eqn3}
\int_{-\infty}^{+\infty}t^m\psi(t)dt=0, ~~~\forall m,~~0\leq m< N
\end{equation}
for example, the wavelet in Figure~\ref{jaman:fig1} has four vanishing moments, i.e. $N$=4.

\subsection{Singularity Detection}

Since the wavelet has $N$ vanishing moments, so~$ \int_{-\infty}^{+\infty}P_n(t-t_0)\psi(t)=0$, (if $n<N$) ,and therefore, the wavelet coefficient only detects the singular part of the signal.
\begin{equation}
\label{jaman:eqn4}
Wf(s,b)\sim s^{\alpha(t_0)},~~~a\rightarrow0^+,
\end{equation}
So, as long as, $N>\alpha(t_0)$ the H\"older exponents \index{H\"older exponent} can be extracted from log-log plot of the Equation~\ref{jaman:eqn4} .
\subsection{Wavelet Transform Modulus Maxima\index{Wavelet Transform Modulus Maxima}}
\label{subsec:WTMM}
Let $[u_p(s)]_{p\in Z}$ be the position of all maxima of $|Wf(b,s)|$ at a fixed scale $s$.
Then the partition function Z is defined as \cite{jaman:mallat}
\begin{equation}
\label{jaman:eqn5}
Z(q,s)=\sum_p|Wf(b,s)|^q
\end{equation}
$Z$ will be calculated from the WTMM. Drawing an analogy from thermodynamics, one can define the exponent $\tau(q)$ from the power law behavior of the partition function~\cite{jaman:Struzik,jaman:ArneodoDNA,jaman:mallat} as
\begin{equation}
\label{jamanjaman:eqn6}
Z(q,s)\sim a^{\tau(q)},~~~~a\rightarrow 0^+,
\end{equation}
The log-log plot of Eqn~\ref{jamanjaman:eqn6} will give the $\tau(q)$ of the signal.

Now the multifractal spectrum $D(\alpha(q))$ vs $\alpha(q)$ can be computed from the Legendre transform
\begin{equation}
\label{jamanjaman:eqn7}
D(\alpha)=min_{_q}(q\alpha-\tau(q))
\end{equation}
where, the H\"older exponent\index{H\"older exponent} $\alpha=\frac{\partial\tau(q)}{\partial q}$. 
\begin{table}
\center
\caption{In this table the fractal dimension\index{fractal dimension} (3-rd column), minimum and maximum (4-th and 5-th column) values of the singularities have been shown for different stations according to their distance from the epicentral distance of the earthquake\index{earthquake}.}
\label{jaman:table1}       
 \begin{tabular}{l|r|r|r|r}
\hline\hline\noalign{\smallskip}
\emph{Earthquake \index{earthquake}recording Station} & \emph{Epicentral Distance(Km)} & Fractal Dim & \multicolumn{2}{c}{Singularity($\alpha$)} \\
    &  & & $\alpha_{max}$    &$\alpha_{min}$\\
\noalign{\smallskip}\hline\noalign{\smallskip}
Santiago de Maria          &  52.50648  & 0.81 & 2.23  & 1.46 \\
Presa 15 De Septiembre Dam &  63.85000  & 0.83 & 2.85  & 1.14 \\
San Miguel                 &  69.95400  & 0.88 & 2.85  & 1.68 \\
 Sensuntepeque             &  90.50100  & 0.84 & 2.53  & 1.40 \\
Observatorio               &  91.02397  & 0.82 & 2.76  & 1.52 \\
Cutuco                     &  96.63410  & 0.84 & 2.60  & 1.48 \\
Santa Tecia                &  97.99589  & 0.91 & 2.58  & 1.69 \\
Acajutia Cepa              & 139.41800  & 0.89 & 3.05  & 1.83 \\
 Santa Ana                 & 142.01300  & 0.86 & 3.32  & 1.49 \\
Ahuachapan                 & 157.35800  & 0.75 & 2.48  & 1.68 \\
Cessa Metapan              & 165.78460  & 0.93 & 3.44  & 1.58 \\
\noalign{\smallskip}\hline
\end{tabular}
\end{table}
\begin{figure}
\center
\includegraphics[height=4cm]{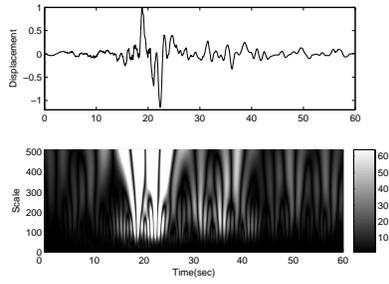}\\
\caption{Typical raw data recorded at station Santa Tecia (Top) and its Continuous wavelet transform(cwt)(bottom). From cwt it is clear that the major earthquake\index{earthquake} events occurs within few seconds (in between 15-20 sec).}
\label{jaman:fig2}
\end{figure}
\begin{figure}  
\center            
\includegraphics[height=4cm]{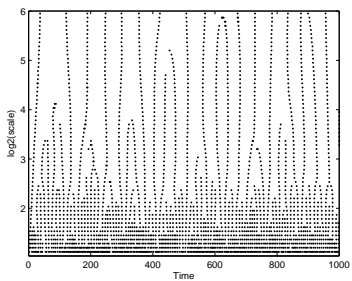}\\
\caption{WTMM skeleton of the data taken at Santa Tacia station(Raw data Fig~\ref{jaman:fig2}).}
\label{jaman:fig3}
\end{figure}
\begin{figure}[h]
\center
\includegraphics[height=4cm]{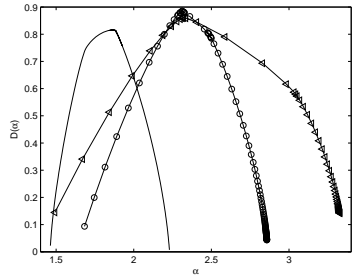}
\caption{Multifractal analysis analysis of El Salvador\index{El Salvador} earthquake\index{earthquake}. In the above figure  $-$, $-0-$, and $-\triangleleft-$, are the singularity spectrums of the data recorded 
at the stations Santiago de Maria, Santa Tacia, and Santa Ana respectively.} 
\label{jaman:fig4}
\end{figure} 
\begin{figure}[h]
\center
\includegraphics[height=4cm]{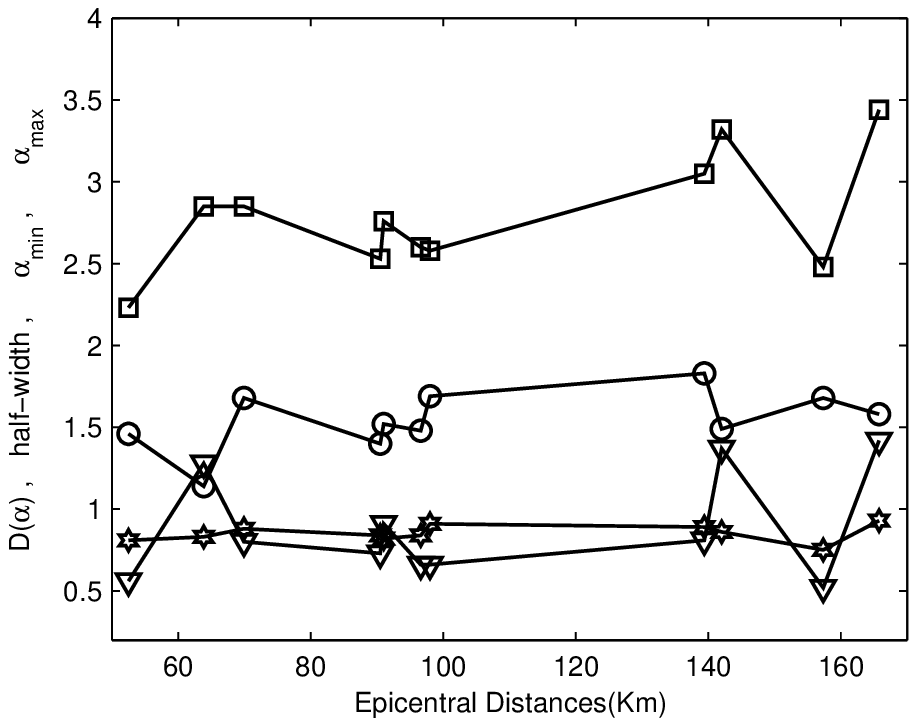}
\caption{From this figure it is clear that the fractal dimension\index{fractal dimension} of singularity support is around 1 ($-*-$), 
lower bound and upper bound of singularity increases with the epicentral distances($-0-$,$-\Box-$ and  respectively)
, and half-width has not such incremental behavior($-\nabla-$). }  
\label{jaman:fig5}
\end{figure} 
\section{Results and Discussion}
\label{ref:Result}  
 In the present paper we have analyzed the El Salvador \index{El Salvador} earthquake\index{earthquake} data recorded at different stations as shown in the Table~\ref{jaman:table1}. In this table we have arranged the stations according to their distances from the epicenter. 

Wavelet analysis of the data recorded at different stations shows that the major events of the earthquake\index{earthquake} have taken place at short time scales. For eg. Fig~\ref{jaman:fig2}(top) shows a burt of activity in a short duration and the corresponding Continuous Wavelet Transform (CWT) in Fig~\ref{jaman:fig2}(bottom) for the time series recorded at Santa Tacia station.  
In this figure(Fig~\ref{jaman:fig2}[bottom]) the maximum correlation is shown by white color (which indicates maximum correlation), which occurs between 15 to 25 seconds approximately shown in fig~\ref{jaman:fig2}. CWT of the recorded data also shows that pseudo frequencies of the major 
events are less than 2 Hz. For Santa Tacia data it is few hundred mHz to 2 Hz. From the same 
figure(Fig~\ref{jaman:fig2}[bottom]) it is also clear that the high frequencies i.e. 1-2 Hz come in very short range (1-4 seconds),  and mHz frequencies comes with relatively longer durations (about 10 seconds).
 Multifractal analysis of the earthquake\index{earthquake} data recorded at different 
 stations of increasing distances from the El Salvador\index{El Salvador} earthquake\index{earthquake} epicenter of 2001 has been carried  out. In the table
 ~\ref{jaman:table1} the first column represents the station according to their distance from the 
 earthquake\index{earthquake}  epicenter (distances shown in the second column is in km). In order to get the multifractal spectrum
 we first calculated the WTMM tree shown in the figure~\ref{jaman:fig3} as described in subsection~\ref{subsec:WTMM}. Using Legendre transform method we have obtained the multifractal 
spectrum shown in the figure~\ref{jaman:fig4}. From multifractal analysis it is clear that the fractal dimension\index{fractal dimension} of the singularity support is around one.
 Lower bound and upper bound of the singularity increases with the distances of the station from the earthquake\index{earthquake} epicenter shown in table~\ref{jaman:table1} and in figure~\ref{jaman:fig4}. It indicates the signal becomes smoother with distance, but the half width of the singularity support has random variation with distances.

In conclusion, the  data shows a multifractal behavior, and the major event takes place in a short duration.

\section*{Acknowledgment}
 Some of the MATLAB function of Wavelab has been used in this analysis( address: http://www-stat.stanford.edu/$\sim$wavelab). 

\bibliographystyle{}
 \bibliography{}
%

\printindex
\end{document}